\input epsf

\documentclass{aa}

\usepackage{babel}

\usepackage[cp850]{inputenc} % remplace ibmkbd
\usepackage[T1]{fontenc}

\usepackage[cp850]{inputenc}
\usepackage[T1]{fontenc}
\usepackage{amsmath,amssymb}
\usepackage[dvips]{graphicx}

\begin{document}

\thesaurus{}
\title{On the possible existence of a self-regulating hydrodynamical
process in slowly rotating stars:}
\subtitle{ I. Setting the stage}
\author{Sylvie Vauclair
              \inst{}
                     }
\titlerunning{Slowly rotating stars}
\authorrunning{Sylvie Vauclair}
\offprints{Sylvie Vauclair}

\institute{ Laboratoire d'Astrophysique, 14 av. Ed. Belin, \\
               31400 Toulouse, France
               }

\date{Received March, 1999; accepted }

\maketitle

\begin{abstract}
It has been known for a long time (Mestel~1953) that the meridional
circulation velocity in stars, in the presence of 
$\displaystyle \mu $-gradients, is the sum of two terms, one due to
the classical thermal imbalance ($\displaystyle \Omega$-currents)
and the other one due to the induced horizontal
$\displaystyle \mu $-gradients ($\displaystyle \mu $-induced
currents, or $\displaystyle \mu $-currents in short). In the most
general cases, $\displaystyle \mu $-currents are opposite to
$\displaystyle \Omega$-currents. Simple expressions for these
currents are derived under some simplifying physical assumptions 
presented in the text, and their physical interpretations are discussed.
Computations of the $\displaystyle \Omega$ and
$\displaystyle \mu $-currents in a 0.8~M$\displaystyle _{\odot}$
halo stellar model including classical element settling show that
the $\displaystyle \mu $-currents are larger (in absolute values)
than the $\displaystyle \Omega$-currents in all the star: some
new physics has thus to be invoked in this case.

We show here how such processes could possibly lead to a quasi-equilibrium
stage in which both the circulation and the helium settling could be cancelled
out. As lithium diffuses in the same way as helium, we expect a very small
lithium concentration gradient below the convective zone in
``plateau stars'' (main-sequence Pop~II stars), much smaller
 than the one expected for pure element
settling. 
This could possibly account for the very small dispersion observed for the
lithium abundances at the surface of these stars. This should also have
important consequences in other contexts which will be discussed
in forthcoming papers. 

The present computations show that
element settling in slowly rotating stars leads to surface abundances
which depend on the competition between $\displaystyle \mu $-currents
and $\displaystyle \Omega$-currents, 
in a way which had not been taken into
account in previous computations. This may change our general
understanding of the diffusion processes of chemical species
in rotating stars.
\end{abstract}

\section{Introduction}

The present study was motivated by the so-called ``lithium plateau
paradox'' (Vauclair 1999), which may be formulated in the following
way:

\begin{itemize}
\item[-] From spectroscopic observations, the lithium abundance in main
sequence Pop~II field stars with effective temperatures larger than
5500~K is remarkably constant. Since the first observations of the
lithium plateau by Spite and Spite (1982), many abundance determination
have confirmed the constancy of the lithium abundance in most of these
stars,
although a small slope may exist as a function of effective temperature
(Thorburn 1994, Ryan et al. 1996, Spite et al. 1996, Bonifacio and Molaro
1997).
 Moreover the dispersion around the average value is extremely
small.
Whether this dispersion is real or below the observational errors is
still a subject of discussion.
(Deliyannis et al 1993, Bonifacio and Molaro 1997, Molaro 1999).
In globular clusters however, some dispersion exists for subgiants close
to the turn-off (Deliyannis et al. 1995, Boesgaard et al. 1998).

\item[-] On the other hand, theoretical computations lead to predictions
of large variations of the lithium abundance from star to star, 
unless some ``ad hoc''
process forces the lithium abundance to remain constant 
(Charbonnel and Vauclair 1998, and references therein). The reason for these predicted
variations is related to the diffusion processes which take place in the
radiative regions inside the stars, below the outer convective zones.
Due to pressure and temperature gradients, lithium settles down towards
the stellar center, as well as helium and other heavier elements.
Mixing induced by rotation, internal waves, or mass loss related
motions may slow down the settling process, but then it brings up to the
convective zone matter in which lithium has been destroyed by nuclear
reactions.

\end{itemize}

These lithium depleting processes have been extensively studied in the
litterature: (e.g. Michaud et al. 1984, Vauclair 1988, Pinsonneault et al.
1992, Vauclair and Charbonnel 1995 and 1998). 
The results are well known and can be summarized as follows:

\begin{itemize}

\item[-] If the radiative zones were completely stable, lithium would
be depleted in Pop~II stars due to settling, and the effect would increase
with effective temperature. As a consequence, 
 the lithium abundances
should decrease with increasing 
$\displaystyle T_{{\rm eff}}$
 in contradiction with the observations. This result
suggests that some macroscopic motion must occur and compete
with the lithium settling. (e.g. Michaud et al.1984, 
Vauclair 1988, Chaboyer
et al. 1992).

\item[-] Rotation-induced mixing, as prescribed by Zahn (1992), or
Pinsonneault et al. (1992), could account for the observations. However
the small dispersion of the observed abundances is difficult to interpret
in this framework (Vauclair and Charbonnel 95) and gives in any case
an upper limit on the initial lithium value (Pinsonneault et al. 1998).

\item[-] Mass-loss could also be a possible explanation (Vauclair
and Charbonnel 1995). However all the stars should have suffer exactly
the same mass loss rate, about 30 times the solar wind rate, during all
their lives.
\end{itemize}

In all cases, the lithium plateau can be accounted for only with
``ad hoc'' hypothesis, where some parameter is assumed constant
(or with very small variations)
in all stars (mass loss rate, rotation rate...). This is not satisfying, and
observers often prefer to forget about this physics and decide that the
lithium abundance observed in Pop~II stars must be the primordial one.

This interpretation however is in contradiction with the strong improvements
of stellar physics obtained these last few years.  The theory of stellar
structure and evolution includes the best available plasma physics, with
new equations of states, opacities, nuclear reaction rates and element
settling. Helioseismology represents an excellent tool for testing this
physics: the agreement of the sound velocity in the models and in the
``seismic Sun'' (deduced from helioseismic modes) is much better
when element settling is introduced (e.g.  Bahcall and
Pinsonneault 1995, Richard et al. 1996,
Brun et al. 1998,  Vauclair 1998).

This is the reason why the lithium plateau becomes a paradox, with
observations in contradiction with the predictions of the best available
theory. What does the  lithium plateau want to tell us that we have
not yet understood ?

The clue may be related to the physics of rotation-induced mixing in the
presence of $\displaystyle \mu $-gradients. This
has been studied by several authors, beginning with Mestel (1953, 1957,
1961, 1965) and Kippenhahn (1958, 1963), and recently by Zahn
(1992) and Maeder and Zahn (1998). In the present paper, the physics
of meridional circulation in the presence of $\displaystyle \mu $-gradients
is revisited. We show that, when one takes into account helium gravitational
settling, $\displaystyle \mu $-induced currents 
opposing the traditional
meridional currents 
(hereafter called $\displaystyle \Omega$-currents)
take place, and that they can become of the same
order of magnitude. Then the circulation freezes out. Furthermore, in this
case the element settling itself may also be prevented. We are in the
presence of a self-regulating process where the rotational currents and
the 
helium settling cancel each other. As lithium diffuses in
a way similar to helium (at the same rate), the lithium settling is
also cancelled by this process. We show how this  could
account for the small dispersion of the lithium plateau. It may also
have other consequences which will be studied in forthcoming
papers.

In section 2, the theory of meridional circulation including the effects
of $\displaystyle \mu $-gradients is recalled. Orders of magnitude
for the case of Pop~II plateau stars are discussed in section~3 
and
evidences of the importance of $\displaystyle \mu $-induced currents
are given. We show how element settling can be prevented when
$\displaystyle \mu $-induced currents
cancel $\displaystyle \Omega$-currents. A general discussion is
presented in section~4.

\section{Meridional circulation including $\displaystyle \mu $-induced currents}
\subsection{Classical meridional circulation}
In rotating stars, the equipotentials of  ``effective gravity'' (including the centrifugal
acceleration) have
ellipso\"{\i}dal shapes while the energy transport still occurs in a spherically
symetrical way. The resulting thermal imbalance must be compensated
by macroscopic motions: the so-called ``meridional circulation''
(Von Zeipel 1924). The stellar regions outside the convective
zones cannot be in complete radiative equilibrium. They
are subject to entropy variations given by (see the review by Zahn 1993):
\begin{eqnarray}
\rho T \left( {\partial S \over \partial t} + {\bf u} \cdot {\bf \nabla} = S\right)
 & = & - {\bf \nabla} \cdot {\bf F} +
\rho  \varepsilon _{n} \nonumber \\
& = & \rho  \varepsilon _{\Omega} \ (\not= 0)
\end{eqnarray}
where $\displaystyle {\bf F}$ represents the heat flux, $\displaystyle \varepsilon _{n}$ the
nuclear 
energy production and $\displaystyle \varepsilon _{\Omega}$ an energy generation
rate which results from sources and sinks of energy along the
 equipotentials.

The vertical component of the meridional velocity $\displaystyle u_{r}$ is computed
as a function of $\displaystyle \varepsilon _{\Omega}$ in the stationary
regime (from eq. 1):
\begin{equation}
u_{r} = \left( {P \over C_{p}\rho T }\right)
{\varepsilon _{\Omega} \over g }
\end{equation}
which, for a perfect gas, reduces to:
\begin{equation}
u_{r} =
{\varepsilon _{\Omega} \over g} \
{{\bf \nabla} _{{\rm ad}}
 \over{\bf \nabla}_{{\rm ad}} - {\bf \nabla} + \nabla_{\mu }}
\end{equation}
where $\displaystyle g$ represents the local gravity, 
$\displaystyle {\bf \nabla}_{{\rm ad}}$ and 
$\displaystyle {\bf \nabla}$ the usual adiabatic and real ratios
$\displaystyle \left( {d \ln T\over d \ln P }\right)$
and $\displaystyle \nabla_{\mu }$ the mean molecular weight
contribution 
$\displaystyle \left( {d \ln \mu  \over d \ln P}
\right)$.

The expression of $\displaystyle \varepsilon _{\Omega}$ is 
computed by expanding the right-hand-side of eq.~(1) on a level surface
and writing that its mean value vanishes.

For nearly-uniform rotation and negligible $\displaystyle \mu $-gradients, one finds
(cf~Mestel 1965, Zahn 1993 and references therein):
\begin{equation}
\varepsilon _{\Omega} =
{8 \over 3 } \ {L \over M }
\left( {\Omega^{2} r^{3} \over GM }\right)
\left( 1 - {\Omega^{2} \over 2 \pi  G\rho }\right)
P_{2} (\cos \theta)
\end{equation}
all quantities being computed at radius $\displaystyle r$.

In this case, neglecting $\displaystyle \nabla_{\mu }$ in (3), 
the vertical meridional velocity may be written:
\begin{equation}
u_{r}=
{8 \over3 } \
{L \over M} \
{\nabla_{{\rm ad}} \over \nabla_{{\rm ad}} - \nabla } \
{\Omega^{2} r \over g^{2} }
\left(1 - {\Omega^{2} \over 2 \pi  G\rho  }\right)
P_{2} (\cos \theta)
\end{equation}
Introducing $\displaystyle U_{r}$ as:
\begin{equation}
u_{r} =
U_{r} \ P_{2} \ (\cos \theta)
\end{equation}
the horizontal meridional velocity is given by (see Mestel 1965):
\begin{equation}
u_{\theta} =
- {1 \over  2 \rho r} \
{d \over dr}\
(\rho r^{2} U_{r})
\sin \theta \cos \theta
\end{equation}

In the presence of mean molecular weight gradients, these
expressions must be modified, as discussed below.

\subsection{$\displaystyle \mu $-induced currents}

Mestel (1953) and (1957) (see also Mestel 1965)
pointed out that, in the presence of vertical $\displaystyle \mu $-gradients,
$\displaystyle \varepsilon _{\Omega}$ contains other terms related
to the resulting horizontal variations of $\displaystyle \mu $:
the so-called ``$\displaystyle \mu $-induced currents''.
The expression of $\displaystyle \varepsilon _{\Omega}$ obtained
in this case, with the assumption of perfect gas law, has been
derived in detail.

More recently Maeder and Zahn (1998),
 hereafter MZ98, gave
a modified expression for 
$\displaystyle \varepsilon _{\Omega}$
which takes into account several effects which were not
included in the previous computations: more general equations
of state instead of perfect gas law, presence of a thermal
flux induced by horizontal turbulence, 
non-stationary cases.

Here we wish to focus on the importance of 
non-negligible
$\displaystyle \mu $-gradients for the meridional
circulation, which may modify our understanding of the 
low-rotation regimes in the presence of microscopic
diffusion.

The basic idea is that on a level surface (with constant
pressure) specific quantities like temperature T, density
$\displaystyle \rho $, gravity $\displaystyle g$,
thermal conductivity $\displaystyle \chi$, rotation
rate $\displaystyle \Omega$ and mean molecular
weight $\displaystyle \mu $ may vary around an
average value. Following 
Zahn 1992 and MZ98 these quantities $\displaystyle x$
are expanded on isobars as:
\begin{equation}
x (P, \theta) = \overline x (P) +
\tilde{x} (P) P_{2}  (\cos \theta)
\end{equation}
The expansion of the energy term in eq.~(1) leads to derivatives in
$\displaystyle {d \over dP }$ which have
to be transformed through an equation of state to lead to
the radius derivatives $\displaystyle {d \over dr }$.

In this procedure, two kinds of terms finally appear:

1) those which are 
directly related to the rotation rate, for example
the gravity fluctuations

2) those which are related to the local 
$\displaystyle \mu $-gradient.

As we discuss a basic physical point, we prefer to simplify
the expressions under specific assumptions, neglecting for
the moment the secondary terms. We assume perfect gas
law, and nearly rigid rotation
(the importance of this assumption will be discussed in
section~4). In this case, the gravity fluctuations are given
by : 
\begin{equation}
{ \tilde g \over g} = { 4\over 3}
\left( {\Omega ^{2}r^{3}\over GM}\right)
\end{equation}

 As we focus on the physical
processes which occur below the convection zones in cool
stars, we also neglect all the terms related to energy production
(stellar cores will be discussed in a forthcoming paper).
Keeping only the non-negligible terms in MZ98's
expressions, we obtain:
\begin{equation}
\varepsilon _{\Omega} =
\left( {L \over M}\right)
\left( E_{\Omega} + E_{\mu }\right) P_{2}
(\cos \theta)
\end{equation}
with:
\begin{eqnarray}{}
E_{\Omega} & = & {8 \over 3}
\left({\Omega^{2}r^{3} \over GM }\right)
\left( 1 - {\Omega^{2} \over 2\pi G\overline \rho  }
\right) \\
E_{\mu } & = &  {\rho _{m} \over \overline \rho  } 
\left\{
{ r \over 3 } \
{d \over dr }
\left[
\left(
H_{T}
{d \Lambda \over dr}\right)
- (\chi_{\mu } + \chi_{T} + 1) \Lambda \right]
- {2 H_{T} \Lambda \over r } \right\}
\end{eqnarray}
Here $\displaystyle \overline \rho $ represents
the density average on the level surface 
$\displaystyle (\simeq \rho )$ while
$\displaystyle \rho _{m}$ is the mean density inside
the sphere of radius $\displaystyle r$; 
$\displaystyle H_{T}$ is the 
temperature scale height;
$\displaystyle \Lambda$  represents the 
horizontal $\displaystyle \mu $ fluctuations
$\displaystyle {\tilde{ \mu}\over \overline \mu } $;
$\displaystyle \chi _{\mu }$ and
$\displaystyle \chi _{T}$ represent the
derivatives:
\begin{equation}
\chi_{\mu } =
\left(
{\partial \ln \chi \over \partial \ln \mu  }\right)_{P,T}
\quad  ; \quad 
\chi_{T} =
\left( {\partial \ln \chi \over \partial \ln T }\right)_{P, \mu }
\end{equation}

In the following we will also neglect the ``Gratton-\"Opik''
term
$\displaystyle {\Omega^{2} \over 2\pi G\rho  }$, which
is justified below the convective zones in slowly-rotating
cool stars.

\subsection{Physical interpretation}

In the presence of $\displaystyle \mu $-gradients, the
circulation velocity is the sum of two terms: 
$\displaystyle E_{\Omega}$,
which we shall call the ``$\displaystyle \Omega$-current'',
and $\displaystyle E_{\mu }$, which we shall call
the ``$\displaystyle \mu $-induced current'' or, for short,
the ``$\displaystyle \mu $-current''.
 In the general case,
$\displaystyle E_{\mu }$ may also be related to $\displaystyle \Omega$
through the horizontal $\displaystyle \mu $-fluctuations
$\displaystyle \Lambda$. We will see however that,
in the stationary case, $\displaystyle \Lambda$ is
independent of $\displaystyle \Omega$. In most of
the star, $\displaystyle E_{\Omega}$ is positive (except
in the very outer layers) while $\displaystyle E_{\mu }$ is
negative (except in case of strong second derivatives of
$\displaystyle \Lambda$,
which are supposed negligible in the following): 
the $\displaystyle \mu $-currents
are opposite to the $\displaystyle \Omega$-currents.
When $\displaystyle \Lambda$ varies
directly like $\displaystyle r$
(this situation occurs in the stationary case, as will
be seen below), $\displaystyle E_{\mu }$ becomes:
\begin{equation}
E_{\mu } =
- 2 \ {\rho _{m} \over \rho  } \
{H_{T} \over r } \Lambda
\left[
1 + {(1 + \chi_{T} + \chi_{\mu }) \over 6 } \
{r \over H_{T} }\right]
\end{equation}
Which we will write:
\begin{equation}
E_{\Omega} \simeq
{8 \over 3 } \
{\Omega^{2} r^{3}\over GM }
\end{equation}
\begin{equation}
E_{\mu } \simeq
- c_{\Lambda} \
{2\rho _{m} \over \rho  } \
{H_{T} \over r } \Lambda
\end{equation}
where 
$\displaystyle c_{\Lambda }$ is of order unity.

The physical interpretation of these currents can be given in
the following hand-waving way. Due to  centrifugal
effects, the effective gravity is constant on ellipsoidal shells
(Figure 1) while the temperature gradients keep the
spherical symetry. Consequently the temperature is larger
at the poles than at the equator on a level surface, leading
to the $\displaystyle \Omega$-currents. The element
abundances are also expected to vary along level surfaces,
as they depend on temperature (in case of gravitational
settling, they also depend on gravity, so that they should
not be constant on spheroids either).

The
density fluctuations on level surfaces are given by (Zahn~1992):
\begin{equation}
{\tilde \rho  \over \rho  } =
{1 \over 3} \ {r^{2} \over\overline g }\
{d\Omega^{2} \over dr }
\end{equation}
As a consequence, in the approximation of nearly rigid rotation, level
surfaces are both surfaces of equal pressure and density.
In this case $\displaystyle \mu $ varies like $\displaystyle T$.

In first approximation, all other parameters assumed constant,
the specific entropy is proportional to the number of
particules per unit volume (Sackur-Tetrode formula, see
Vauclair~1993), or inversely proportional to $\displaystyle \mu $.
Thus the horizontal $\displaystyle \mu $-gradients
lead to a larger entropy at the equator than at the pole, which
explains why $\displaystyle \mu $-currents are opposite
to $\displaystyle \Omega$-currents.

%\begin{figure}[p] % met les figures a la fin du texte
\begin{figure}
\vbox{\epsfxsize=9cm \epsfbox{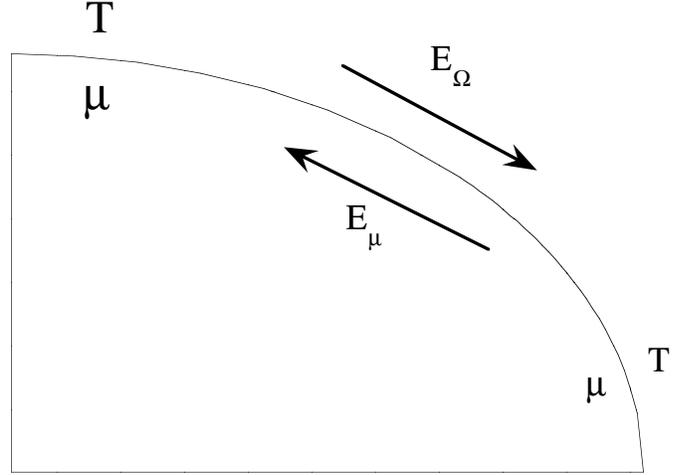}}
%\vspace{5cm}
\caption{Schematic drawing of a ``level surface'' in a rotating
star. Temperature and mean molecular weight are larger at the poles
than at the equator, inducing a competition between two effects: the
so-called ``$\displaystyle \Omega$-currents'' from pole to equator
and ``$\displaystyle \mu $-currents'' from equator to pole. At
equilibrium, the circulation stops. Then element settling proceeds at
the pole in a more rapid way than at the equator, thereby decreasing
the horizontal $\displaystyle \mu $-gradient and
$\displaystyle E\mu $. The circulation begins again with
matter going up at the pole: it brings back the material in which helium
has settled, restablishing the equilibrium horizontal 
$\displaystyle \mu $-gradient. Such a process can cancel altogether
meridional circulation and helium settling.
}
\end{figure}

The $\displaystyle \mu $-currents induced by the
flattening of the level surfaces are very small though, so that
they do not prevent the onset of the meridional circulation.
However, as soon as the circulation begins, the horizontal
$\displaystyle \mu $-gradients increase as the upward
flows bring $\displaystyle \mu $-enriched matter
at the poles while the downward flows bring
$\displaystyle \mu $-depleted matter at the equator.
As will be seen below, a stationary solution can occur
for which, in some cases, the $\displaystyle \mu $-currents
may become of the same order as the $\displaystyle \Omega$-currents,
leading to new interesting physics.

\subsection{Discussion on the horizontal
 variations of $\displaystyle \mu $}

The order of magnitude of the
 horizontal $\displaystyle \mu $-gradients
induced by the flattening of the level surfaces 
$\displaystyle \Lambda_{f}$ can be
obtained with the approximation that the radius differences
between the polar and equatorial regions are given by:
\begin{equation}
\lambda  =
{\Delta r \over r } \simeq
{\Omega^{2} r^{3} \over  GM}
\end{equation}
The values of this flattening parameter are given in Figure~2 as a
function of $\displaystyle r$
in a $\displaystyle 0.8$~M$\displaystyle _{\odot}$
halo star which rotates with a velocity of
$\displaystyle 10$~km.s$\displaystyle ^{-1}$.

In this case, the horizontal $\displaystyle \mu $ fluctuations can
be written:
\begin{equation}
\Lambda_{f } =
{\tilde \mu  \over \mu  } \simeq
{\Omega^{2} r^{4} \over GM } \cdot \nabla \ln \mu 
\end{equation}
or, with the following definition of the ``$\displaystyle \mu $-scale height'':
\begin{equation}
H_{\mu } = (\nabla \ln \mu )^{-1}
\end{equation}
the horizontal fluctuations related to the flattening become:
\begin{equation}
\Lambda_{f } =
{\Omega^{2}r^{3} \over GM  } \
{r \over H_{\mu } }
\end{equation}
These fluctuations are small compared to those
which occur as soon as the meridional circulation is established. In
case of a laminar circulation (pure advection) the order
of magnitude of $\displaystyle \Lambda$ is given by:
\begin{equation}
\Lambda_{a} \simeq
r \cdot \nabla \ln \mu  \simeq
{r \over H_{\mu }}
\end{equation}
According to Zahn (1992) and Chaboyer and Zahn (1992),
the shears induced by the circulation lead to a strong
horizontal turbulence, which reduces the horizontal
$\displaystyle \mu $-gradients.
In this case,
introducing a  horizontal diffusion coefficient
$\displaystyle D_{h}$, the local variations of
$\displaystyle \mu $ are solutions of:\footnote{
Following Chaboyer and Zahn (1992) and MZ 98,
$\displaystyle \mu $ 
is developped on $\displaystyle r$ instead of
$\displaystyle P$. The difference 
is negligible though, as the flattening effect on
$\displaystyle \mu $ is extremely small.}
\begin{equation}
{\partial\tilde \mu  \over \partial t } +
u(r) \cdot
{\partial\overline \mu  \over dr } =
- {6 \over r^{2}}\ D_{h} \ \tilde \mu 
\end{equation}
In the stationary case, we obtain for the horizontal
$\displaystyle \mu $-gradient including turbulence:
\begin{equation}
\Lambda_{t} = {\tilde \mu  \over \overline \mu  } =
- {U_{r} \cdot r^{2} \over 6D_{h} } \
{\partial  \ln \mu \over \partial r }
\end{equation}
With the assumption that
$\displaystyle D_{h}  $ is proportional to
$\displaystyle U_{r} \cdot r$ and not simply to
$\displaystyle U_{r}$ as written by mistake in
MZ98, eq. (22) becomes:
\begin{equation}
\Lambda_{t} \simeq - {1 \over 6\alpha _{h}}\
{\partial \ln \mu  \over \partial \ln r }
\simeq 
- {1 \over 6\alpha _{h}} \
{r \over  H_{\mu } }
\end{equation}
Replacing in eq.~(16) gives the simple expression:
\begin{equation}
E_{\mu } = - {1 \over 3\alpha _{h} } \
{H_{T} \over H_{\mu } } \
{\rho _{m} \over \overline \rho  }
\end{equation}
with $\displaystyle \overline \rho \simeq \rho $
 and
$\displaystyle \rho _{m} = {3 M \over 4\pi  r^{3} }$

\begin{figure}
\vbox{\epsfxsize=9cm \epsfbox{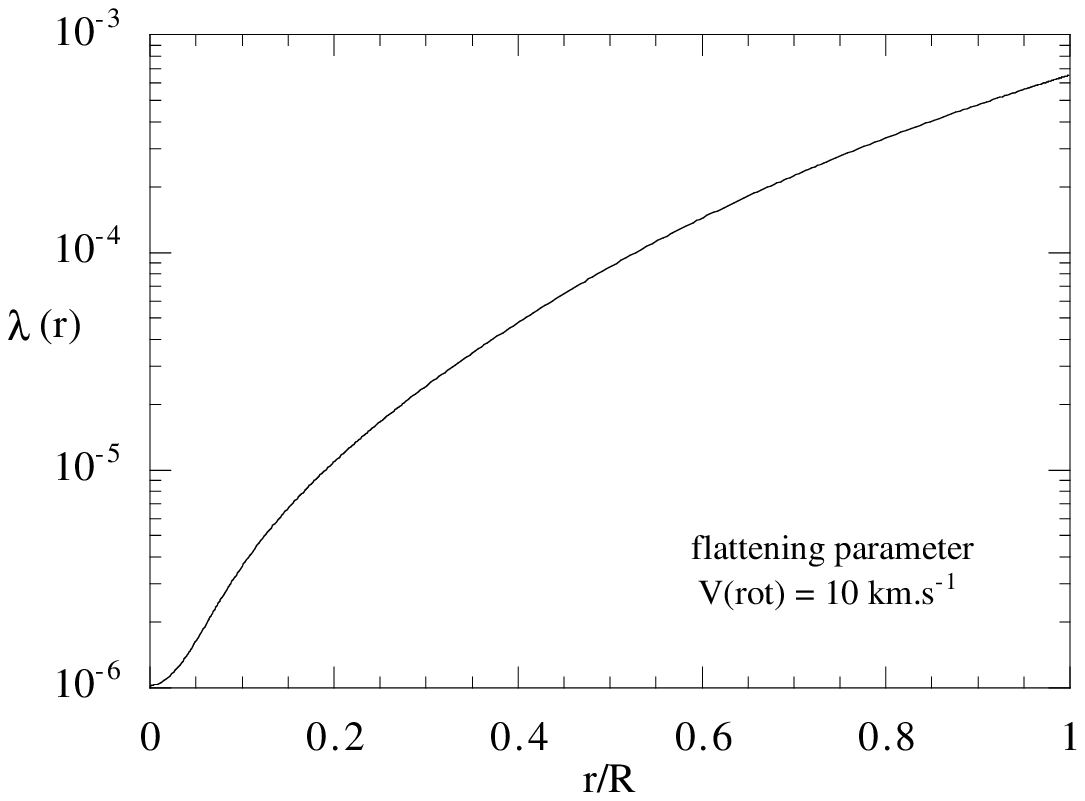}}
\caption{Flattening parameter 
$ \displaystyle\lambda  = {\Delta r \over r} \simeq 
{\Omega^{2}r^{3} \over GM }$ 
in a 0.8~M$\displaystyle _{\odot}$
plateau star with 
$\displaystyle V_{{\rm rot}} = 10$~km.s$\displaystyle ^{-1}$
(see text for details).}
\end{figure}

In summary, 
in the most likely case of meridional circulation associated
with a strong horizontal turbulence,
 the vertical
velocity of meridional circulation can simply be written:
\begin{eqnarray}{}
u_{r} & = & U_{r} P_{2} (\cos \theta)  \nonumber \\
{\rm with} \quad 
U_{r} & = & {1 \over g } \
\left( {{\bf \nabla}_{{\rm ad}} \over 
{\bf \nabla}_{{\rm ad}} - {\bf \nabla} + {\bf \nabla}_{\mu }}\right)
\ \left({L \over M }\right) \
\left(E_{\Omega}  + E_{\mu }\right) \\
E_{\Omega} & = &  {8 \Omega^{2} r^{3} \over 3 GM } \\
E_{\mu } & = & 
- {C_{\Lambda} \over \alpha _{h} } \
{M \over 4 \pi  r^{3} \rho  } \
{H_{T} \over H_{\mu } }
\end{eqnarray}{}
According to Zahn's (1993) prescriptions, 
while the horizontal diffusion
coefficient is larger than
$\displaystyle U_{r} \cdot r$ \ 
$\displaystyle (D_{h} = \alpha _{h} U_{r} \cdot r$
with $\displaystyle \alpha _{h} > 1)$,
the vertical transport may be approximated as an
effective diffusion coefficient of order:
\begin{displaymath}
D_{{\rm eff}} \simeq {1 \over C_{h} }
U_{r} \cdot r \quad \quad  {\rm with} \quad \quad 
C_{h} \simeq 30 
\end{displaymath}

The reason why $\displaystyle E_{\mu }$ does not depend
on $\displaystyle \Omega$
but depends on $\displaystyle H_{\mu }^{-1}$ may be
physically interpreted in the  following way. 
Suppose that at level $\displaystyle r$
the meridional circulation can be approximated by an
upwards and a downwards moving flows. The horizontal
turbulence, described by the horizontal turbulent diffusion
coefficient $\displaystyle D_{h}$, leads to matter transfer
from one flow to the other. In this framework, a vertical
$\displaystyle \mu $-gradient induces a horizontal
$\displaystyle \mu $-gradient which plays a non-negligible
role in the entropy balance and has a feed-back effect on the
velocity $\displaystyle U_{r}$. Increasing the 
rotation velocity should increase $\displaystyle U_{r}$,
the horizontal $\displaystyle \mu $-gradient and the corresponding
$\displaystyle D_{h}$. But the increase of the mixing
effect related to $\displaystyle D_{h}$ decreases
this horizontal $\displaystyle \mu $-gradient.
The horizontal turbulence ``swallows'' the possible variations
induced by different rotation velocities, leaving the vertical mixing
process unchanged.

\begin{figure}
\vbox{\epsfxsize=9cm \epsfbox{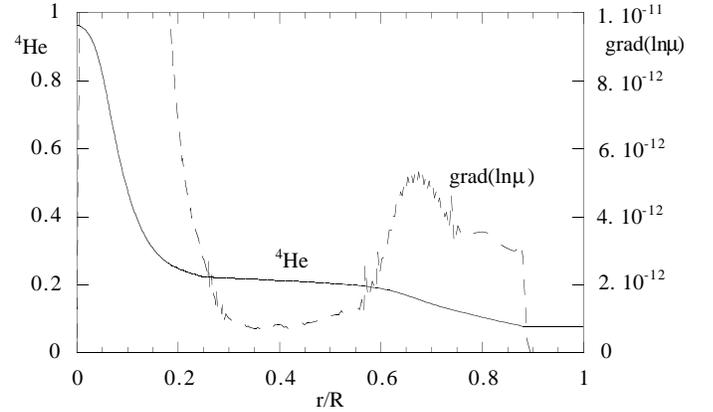}}
\caption{Helium
abundance profile and grad($\displaystyle \ln \mu $) in a
0.8~M$\displaystyle _{\odot}$ plateau star, 
including element settling,
at the age of
12~Gyr. The increase in grad($\displaystyle \ln \mu $) below
the convective zone is clearly related to the $\displaystyle ^{4}$He slope.
}
\end{figure}

\section{Orders of magnitude and discussion}
\subsection{Comparison between the $\displaystyle \Omega$-currents and
the $\displaystyle \mu $-currents}

At this point, it is interesting to compare the orders of magnitudes
of $\displaystyle E_{\Omega}$ and $\displaystyle E_{\mu }$
in stellar models. As the present paper has been motivated by the existence
of the lithium plateau in halo stars, we chose to discuss as an
example the case of a 0.8~M$\displaystyle _{\odot}$ low
metallicity star, situated at the hot end of the plateau
($\displaystyle Z = 10^{-3}$; age: 12~Gyr).
Figure~2 gives the values of the flattening parameter
$\displaystyle \lambda (r) = {\Omega^{2} r^{3} \over GM }$,
with the assumption of a stellar rotation velocity
$\displaystyle V_{{\rm rot}} = 10$~km.s$^{-1}$
which represents an upper limit for these slowly rotating stars.
Figure 3 displays the abundance 
 profile in the same star, including element settling
(Charbonnel and Vauclair 1999). It also shows the
 $\displaystyle \mu $-gradient $\displaystyle \nabla \ln \mu $,
which is essentially related to the helium profile.

\begin{figure}
\vbox{\epsfxsize=9cm \epsfbox{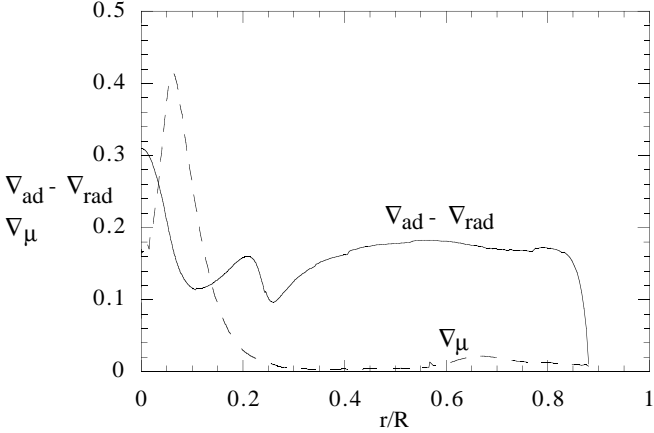}}
\caption{
Comparison of the $\displaystyle \mu $-gradient 
$\displaystyle {d \ln \mu  \over d \ln P }$ with
the usual adiabatic and radiative gradient differences. The
$\displaystyle \mu $-gradient is always negligible
compared to the others,
except in the core and right at the bottom of the convective zone.
}
\end{figure}

The values of 
$\displaystyle \nabla_{\mu } = \left( {\partial \ln \mu 
 \over \partial \ln P }\right)$
are given in Figure~4 also
for the model including element settling,
together with the difference
$\displaystyle \nabla_{{\rm ad}} - \nabla$: we can see
that $\displaystyle \nabla_{\mu }$ is always negligible
except in the central core 
and right at the bottom of the
convective zone. The $\displaystyle \mu $-scale heights
$\displaystyle H_{\mu }$, defined as 
$\displaystyle\left( { d \ln \mu \over dr }\right)^{-1} $ 
are given in
Figure~5 for two models of the same star, one without
element settling (st) and the one including element settling (dif).
The temperature scale height $\displaystyle H_{T}$,
computed in the model with element settling, is also shown
(the difference in
$\displaystyle H_{T}$ between the two models is negligible).
The effect of helium diffusion is
 clearly visible on this graph. 

\begin{figure}
\vbox{\epsfxsize=9cm \epsfbox{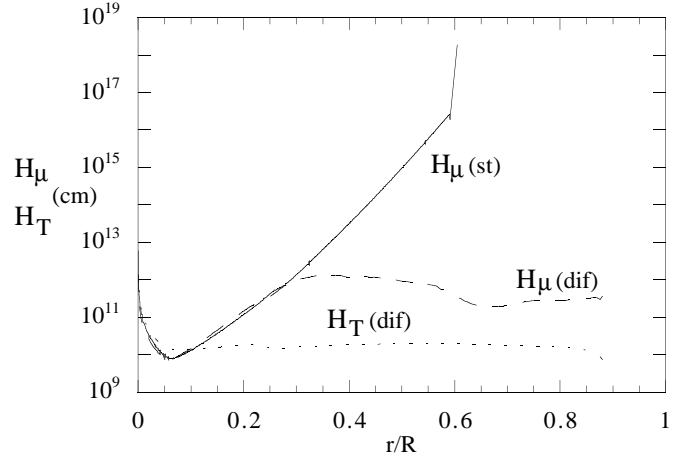}}
\caption{
$\displaystyle \mu $-scale heights computed in two different
0.8~M$\displaystyle _{\odot}$
plateau stars models: one with no element settling included (st)
and one with element settling included (dif). The temperature
scale-height is also shown for the model with element settling
(there is no significant difference with the other model).
The difference in the $\displaystyle \mu $-scale heights
is the signature of helium settling.
}
\end{figure}

Figure 6
presents for these two models a comparison of the
absolute values of $\displaystyle E_{\mu }$ and
$\displaystyle E_{\Omega}$. Here again 
the rotation velocity is taken as
10~km.s$\displaystyle ^{-1}$
which represents an upper limit for halo stars.
The $\displaystyle E_{\mu }$ values are computed
with the assumption of meridional circulation and
horizontal turbulence, and the diffusion coefficient
$\displaystyle D_{h}$ is assumed equal to
$\displaystyle 10 \ U_{r} \cdot r \ (\alpha _{h} = 10)$.
We can see that in the model 
without settling, 
$\displaystyle \vert E_{\mu }\vert$ is larger
than $\displaystyle \vert E_{\Omega} \vert$ in the
central regions and smaller in the outer layers
while in the model including settling, 
 $\displaystyle \vert E_{\mu }\vert$
is always larger than 
$\displaystyle \vert E_{\Omega}\vert$.
This means
that some physical process must occur which has not yet been
included in the theory.
The case of the stellar core, where other effects can occur,
will be discussed in a forthcoming paper. Here we focus
on the radiative region below the convective zone.

Suppose the star begins on the main sequence with homogeneous
abundances. Then $\displaystyle \vert E_{\mu } \vert$
lies below $\displaystyle \vert E_{\Omega} \vert$
and the meridional circulation can occur in the same way
as we have already discussed.
Horizontal turbulence may develop, and a vertical mixing
takes place with an effective diffusion coefficient
$\displaystyle D_{{\rm eff}}$ (Zahn 1992). This 
vertical mixing slows down the element settling without
stopping it. Defining $\displaystyle H_{c}$ as the
concentration scale height, the element settling would
stop only for:
\begin{equation}
{D_{{\rm eff}} \over H_{c} } \simeq
{k_{p} D_{m} \over H_{p} }
\end{equation}
where $\displaystyle D_{m}$ is the microscopic
diffusion for the considered  element and
$\displaystyle k_{p}$ the coefficient
in front of the pressure gradient in the microscopic
diffusion velocity (see, for example, Vauclair and
Vauclair 1982).
However, before reaching this equilibrium concentration,
{\bf the induced $\displaystyle \mu $-currents
become of the same order 
as the $\displaystyle \Omega$-currents},
therefore stopping the circulation.

\begin{figure}
\vbox{\epsfxsize=9cm \epsfbox{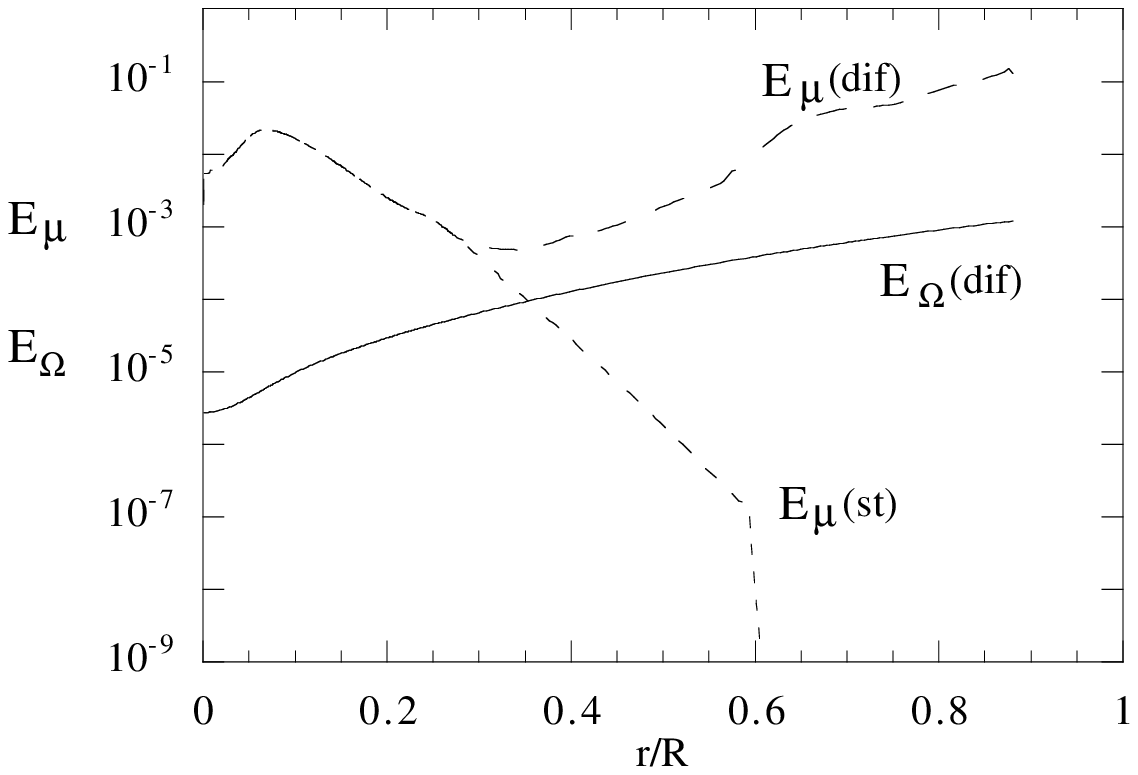}}
\caption{
$\displaystyle \mu $-currents (in absolute values) computed
in the two 0.8~M$\displaystyle _{\odot}$ plateau stars
models, with and without element settling, and 
$\displaystyle \Omega$-current computed in the model
with settling, assuming $\displaystyle V_{{\rm rot}}= 10$~km.s$\displaystyle ^{-1}$
and $\displaystyle \alpha  = 10$ (see text).
We can see that in the model with settling, 
$\displaystyle \vert E_{\mu } \vert$ is always larger
than
$\displaystyle \vert E_{\Omega}\vert$. As most halo stars have
rotation velocities below this value, $\displaystyle \vert E_{\Omega}\vert$
is expected to be even smaller than shown here, which reimforces this conclusion.
}
\end{figure}

We can compute, for the same
 0.8~M$\displaystyle _{\odot}$ star,
the critical $\displaystyle \mu $-scale height for
which 
 $\displaystyle \vert E_{\mu } \vert = \vert
E_{\Omega}\vert$.

We obtain:
\begin{equation}
H^{{\rm crit}}_{\mu } =
{H_{T} \over 8\alpha_{h} \lambda  } \
{\rho _{m} \over \rho  } =
{H_{T} \over \alpha _{h} } \
{3GM^{2} \over 32 \pi  \Omega^{2}\rho r^{6} }
\end{equation}
The profile of 
$\displaystyle H^{{\rm crit}}_{\mu } (r)$
is given in Figure~7, also for
$\displaystyle V_{{\rm rot}} = 10$~km.s$\displaystyle ^{-1}$
and $\displaystyle \alpha _{h} = 10$. For different rotation velocities
$\displaystyle H^{\rm crit}_{\mu }$ is simply multiplied by
$\displaystyle (100/v_{\rm rot})^{2}$.

It is interesting here to compare this critical
$\displaystyle \mu $-scale height with the
concentration scale height given by equation (30). For
a simple mixture of completely ionised hydrogen and
helium, the mean molecular weight is given by:
\begin{equation}
\mu \simeq {1 + 4 c \over 2 + 3c }
\end{equation}
With $\displaystyle c$ of the order of 0.1, we find
for helium:
\begin{equation}
H_{\mu } \simeq 6 H_{c}
\end{equation}
Equation (30) gives, for the equilibrium concentration scale
height:
\begin{equation}
H_{c} =
H_{p} {D_{{\rm eff}} \over k_{p} D_{m} }
\end{equation}
Taking $\displaystyle k_{p} \simeq 5$,
$\displaystyle H_{p} \simeq 10^{10}$ and
$\displaystyle D_{{\rm eff}}/D_{m} \simeq 100$
we find:

$\displaystyle H_{c} \simeq 2 \times 10^{11}$~cm
and $\displaystyle H_{\mu } \simeq 10^{12}$~cm.

We can check from Figure~7 
that the critical $\displaystyle H_{\mu }$
which stops the circulation is reached much before
the equilibrium $\displaystyle H_{c}$
which
would stop the helium settling.

Now the question arises: what happens when
$\displaystyle \vert E_{\mu }\vert$ becomes
equal to $\displaystyle \vert E_{\Omega}\vert$ ?

\begin{figure}
\vbox{\epsfxsize=9cm \epsfbox{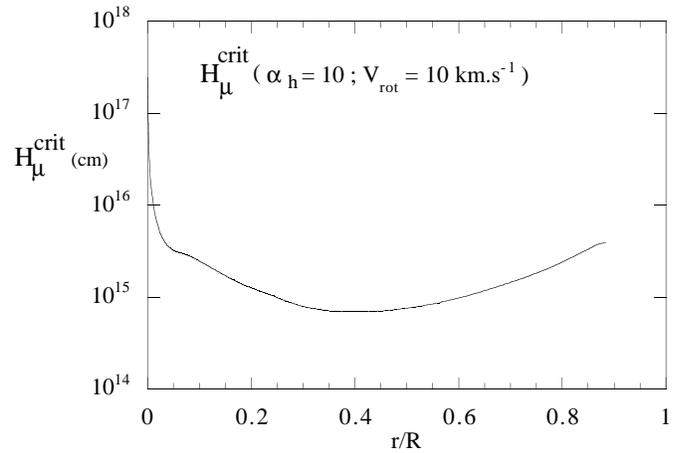}}
\caption{
Critical $\displaystyle \mu $-scale height for which
$\displaystyle\vert E_{\mu }\vert
= \vert E_{\Omega}\vert $ in 
a 0.8~M$\displaystyle _{\odot}$
plateau star, with the
same assumptions as in Figure~6. In this case, the 
$\displaystyle \mu $-scale height is more than 5 orders of
magnitude larger than the stellar radius: the resulting helium (and
also lithium) profiles should be flat below the convection zone. For
lithium, it should remain flat down to the nuclear destruction region.
As $\displaystyle H^{\rm crit}_{\mu }$, varies like 
$\displaystyle (v_{\rm rot})^{-2}$, this conclusion is reimforced
for slower rotations.}
\end{figure}

\subsection{Self-regulating process in case of meridional
circulation and settling}

Let us first summarize the situation of a slowly rotating star
in which element settling leads to an increase of the
$\displaystyle \mu $-gradient below the outer convection
zone.

At the beginning, the star is homogeneous and meridional
circulation can occur, leading to upward flows in the polar
regions and downward flows in the equatorial parts
(except in the very outer layers where the Gratton-\"Opik
term becomes important, which we do not discuss here).
The $\displaystyle \mu $-currents, opposite to the
classical $\displaystyle \Omega$-currents, are first
negligible. The $\displaystyle \mu $-gradients 
increasing with time because of helium settling, the order
of magnitude of the $\displaystyle \mu $-currents also
increase and the $\displaystyle \mu $-scale height
decreases until it reaches $\displaystyle 
H^{{\rm crit}}_{\mu }$ for which the circulation vanishes.

This does not occur all at once:
as $\displaystyle H^{{\rm crit}}_{\mu }$ decreases
with depth below the convective zone (Figure~7) we expect that the meridional
circulation freezes out step by step. An equilibrium situation
may be reached, in which the temperature and mean molecular
weight gradients along the level surfaces are such that
$\displaystyle \Omega$-currents and 
$\displaystyle \mu $-currents cancel each other.

Once it is reached, this equilibrium situation is quite robust. Suppose that some 
mechanism leads to a decrease of the horizontal
 $\displaystyle \mu $-gradient: then $\displaystyle \vert E_{\mu }\vert$ becomes smaller
then $\displaystyle \vert E_{\Omega }\vert $ and the circulation
tends to be restablished in the $\displaystyle \vert E_{\Omega }\vert $
direction. The flow goes up in the polar regions, bringing back matter with
larger $\displaystyle \mu $: the $\displaystyle \mu $ gradient is restored
and the circulation stop.

Suppose now that the horizontal
 $\displaystyle \mu $-gradient is increased.
Then $\displaystyle \vert E_{\mu }\vert$ becomes larger than
$\displaystyle \vert E_{\Omega }\vert $ and the circulation begins in 
the $\displaystyle E_{\mu }$ direction. Matter with larger 
$\displaystyle \mu $ goes up in the equatorial region, decreasing
the $\displaystyle \mu $-gradient. Here again the circulation stops.

As there is no more circulation, we would expect that helium
settling can proceed further.
The microscopic diffusion velocity, in case of pure gravitational
settling, can simply be written (Vauclair and Vauclair 1982):
\begin{equation}
V_{D} = - D_{m}
{k_{p} \over H_{p} } =
- D_{m} k_{p} {\mu g \over kT }
\end{equation}
with:
\begin{equation}
D \ \alpha  \ T^{5/2}/\rho 
\end{equation}
On a level surface,
$\displaystyle \mu $ varies like $\displaystyle T$,
so that
$\displaystyle v_{D} \ \alpha  \ T^{5/2}$
and
$\displaystyle {\Delta v_{D} \over v_{D} } =
{5 \over 2 } \ {\Delta  T\over T } = {5 \over 2} \Lambda$.

As a consequence, on a level surface, the helium gravitational
settling is more efficient
 in the regions of larger $\displaystyle T$,
namely the polar regions, than in the equatorial regions. 

We can now describe the scenario: once the meridional
circulation is frozen, the helium gravitational settling
would like to
proceed and lead to a
decrease of the horizontal $\displaystyle \mu $-gradient.
However the
$\displaystyle \mu $-currents
become  smaller than the 
$\displaystyle \Omega$-currents and
  $\displaystyle \mu $-enriched matter
is brought back into the
polar region, restoring the original $\displaystyle \mu $-gradient.

 We may be here in the presence of a self-regulating process
in which the $\displaystyle \mu $-gradients
which should increase with time
due to the settling are prevented to do so because of  the
currents equilibrium.
Such a process would stop altogether the meridional
circulation {\bf and} the element settling !

In this case the element abundances that we observe at the
surface would be the result of the self-regulating process
and not the result of pure settling. For halo stars, as shown on
Figure~7, this would occur for $\displaystyle H_{\mu }$
much larger than the stellar radius: the abundance profiles would
be flat below the convection zone.

This process could be the reason for the low dispersion of the
lithium abundance in the lithium plateau of halo stars: the
lithium abundance profile would be flat down to the place where
the diffusion time scale is of the order of the nuclear destruction
time scale, that is the place of the lithium peak (Vauclair and
Charbonnel 1998). The observed abundance would then depend
only on the microscopic physical parameters, constant from
star to star, which would explain the very small observed
dispersion. This process will be studied numerically in a forthcoming paper.

\section{Conclusion}
Computations of meridional currents in the presence of $\displaystyle \mu $-gradients show
that, as already pointed out long ago by Mestel (1953), and recently studied
by Zahn (1992) and Maeder and Zahn (1998), $\displaystyle \mu $-induced
currents 
($\displaystyle \mu $-currents in short)
settle in opposition to $\displaystyle \Omega$-currents. In the
present paper, we have computed the orders of magnitude of these
currents under simplifying assumptions. Among these assumptions (energy
production negligible, perfect gas equation of state, Gratton-\"Opik term
negligible, nearly solid rotation), the most stringent one is the assumption
of nearly-solid rotation, as pointed out by Zahn (1999). Assuming solid
rotation means that some other process like internal waves must have
forced the transport of angular momentum (Zahn et al. 1997). In
any case, we know from helioseismology that the Sun does indeed rotate
as a solid body below the outer convective zone (Brown et al. 1989).
This is not yet quite understood, but we may infer that the same process
which forces solid rotation inside the Sun also acts in other solar-type
stars. This may not be crucial for our general conclusions, although
including differential rotation would need a more complicate treatment
of the equations, and would add another unknown parameter. This
may be studied in the future.

We have shown that under these assumptions $\displaystyle \mu $-currents
may become of the same order of magnitude as $\displaystyle \Omega$-currents
in case of element settling, long before the concentration gradient induced
in case of pure settling is reached. In the case of a 
0.8~M$\displaystyle_{\odot} $ halo star with standard
helium settling, 
we have computed that the $\displaystyle \mu $-currents
 would be several orders of magnitude larger
than the $\displaystyle \Omega$-currents ! This is also the case
in the central stellar regions, due to nuclear reactions.

Here we have only focused on the physical processes which occur below
the outer convective zones. In the case of  halo stars we found that, for
slow rotation, $\displaystyle \mu $-currents cancel
$\displaystyle \Omega$-currents for very small concentration gradients,
of order $\displaystyle 10^{-15} (H^{{\rm crit}}_{\mu } = 10^{15}$~cm).
Furthermore, in this case helium setlling may also be cancelled out as it
would decrease the effect of horizontal $\displaystyle \mu $-gradients,
thereby setting again meridional
currents which would restore the original gradients.
This may become a self-regulating process which could explain why the
observed lithium in the plateau stars has such a small dispersion: the
inferred
lithium gradient inside the star, below the convective zone, would
be very small and the lithium profile nearly flat until the
lithium nuclear destruction region.
This
would give a surface lithium abundance close to that of
 the ``lithium peak''
(Vauclair and Charbonnel 1998), the same one for all the stars.

Numerical simulations of the whole process have to be done, to check
all these effects. It will also be tested for the solar case (including central
regions), and for the case of Am stars (for which the Gratton-\"Opik term
is not negligible), in forthcoming papers.
There are many observations in stars which give evidences of mixing processes
occuring below the outer convection zones as, for example, the lithium depletion
observed in the Sun and in galactic clusters. The process we have described
here should not apply in all these stars. The reason could be related to the
rapid rotation of young stars on the ZAMS and to their subsequent
 rotational braking
(here we suppose that the stars always rotated slowly). Also for stars
in which the Gratton-\"Opik term is no more negligible, the whole process
has to be revised.
 Here we have only studied equilibrium situations, while time-evolving
processes should be tested in the future. Finally other mixing processes,
like ABCD or GSF instabilities, could play a role.

In any case the present
computations show that element settling in slowly rotating stars leads to
surface abundances which depend on the competition between
$\displaystyle \mu $-currents and $\displaystyle \Omega $-currents.
This may change our general understanding of the diffusion processes
of chemical species in rotating stars.

\end{document}